\documentclass[journal=ancac3,manuscript=article]{achemso}
\usepackage[version=4]{mhchem}
\usepackage[T1]{fontenc}
\usepackage{textgreek}
\usepackage{amssymb}
\usepackage{epstopdf}
  \DeclareGraphicsExtensions{.png,.pdf}

\author{Vincent Mittag}
\author{Christian Strelow}
\author{Tobias Kipp}
\email{tobias.kipp@uni-hamburg.de}
\phone{+49 40 42838-8277}
\author{Alf Mews}
\affiliation[University of Hamburg]
{Insitute of Physical Chemistry, University of Hamburg, Hamburg, Germany}

\title[CdSe-Dot/CdS-Rod/PbS-Dot Nanocrystals by Partial Cation Exchange Reaction]
  {CdSe-Dot/CdS-Rod/PbS-Dot Nanocrystals by Partial Cation Exchange Reaction}

\begin{document}

\begin{tocentry}
    \includegraphics[scale=0.5]{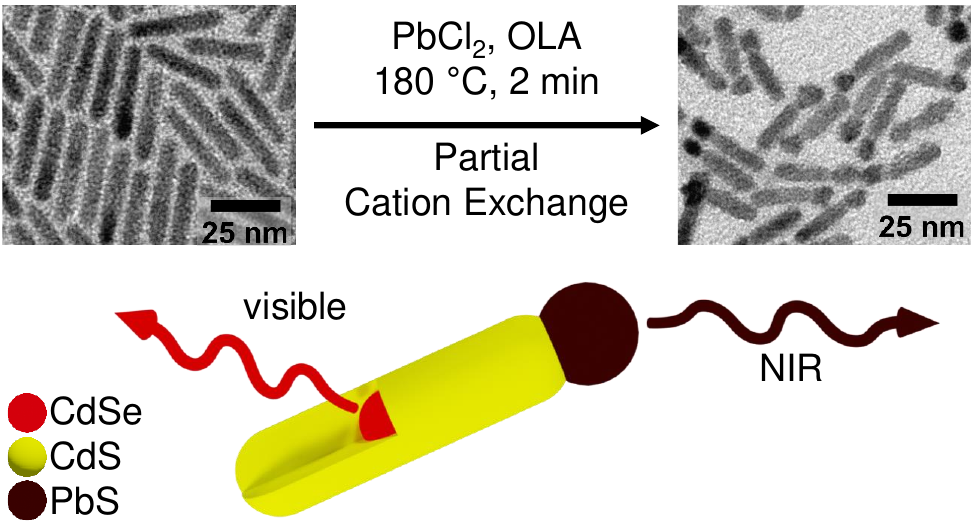}
\end{tocentry}

\begin{abstract}
Dual-emissive nanorods with fluorescence in both, the visible and infrared range are prepared by a combination of a CdSe-nanocrystal-seeded growth of CdS nanorods, and a successive partial Cd-to-Pb cation exchange. 
We show that the exchange reaction, which involves Pb halides in oleylamine, starts at the tip of the rods, leading to the formation of CdSe-dot/CdS-rod/PbS-dot nanocrystals (DRDs). Besides these DRDs, the reaction product also contains shorter nanorods and spherical quantum dots. Their fraction strongly depend on the amount of lead halide  precursor and the reaction time. 
The reaction mechanism is investigated in detail, such that by carefully adjusting the reaction conditions, it is possible to synthesize DRDs of distinct PbS-dot sizes with yields over 95\%.
The resulting DRDs are crystalline and show a CdSe-fluorescence band in the visible range at 600 nm, and also a fluorescence band in the NIR at 1440 nm, resulting from the PbS part of the rods.
  
\end{abstract}


\section{Introduction}

Combining different materials in one nanoparticle leads to heteronanostructures with new and unique properties. For example, rather simple core/shell structures of spherical semiconductor nanocrystals exhibit enhanced photoluminescence quantum yields \cite{Dabbousi1997}, reduced blinking \cite{Chilla2008} or increased stability against photo oxidation\cite{Stouwdam2007}. More complex structures reveal properties like dual emission\cite{Battaglia2005}, intraparticle charge transfer\cite{Enders2018}, two-color anti-bunching\cite{Deutsch2012} or upconversion\cite{Teitelboim2016a}. Recently, two core/shell nanocrystals were coupled and showed the behavior of an artificial molecule.\cite{Cui2019}

The syntheses of heteronanoparticles demand highly controllable reactions and multistep strategies. Wet-chemical techniques for the synthesis of nanoparticles are the hot-injection method and cation-exchange (CE) procedures, for example. In many cases, CE is used to obtain nanostructures, which cannot be synthesized by hot-injection or other methods.\cite{DeTrizio2016,Rivest2013} By a partial CE on a nanoparticle, it is possible to obtain heterostructures like core/shell\cite{Pietryga2008,Justo2014}, Janus\cite{Zhang2015,Lesnyak2015}, or even more complex nanoparticles\cite{Fenton2018}. Whether Janus particles or complex nanoparticles consisting of two or more materials are produced, depends on the anisotropy of the crystal structure\cite{Sadtler2009} and/or the CE\cite{Robinson2007,Zhang2015}. Therefore, CE is a versatile instrument for the synthesis of complex heterostructured nanoparticles.\cite{Beberwyck2013}

In case of the material combination CdSe/PbSe, an important step was done by Zhang et al.\ by reporting on the synthesis of spherical PbSe quantum dots by a direct and full CE from Cd to Pb in spherical CdSe nanocrystals. This was achieved using \ce{PbCl2} in oleylamine as a precursor.\cite{Zhang2014} By refining the method, they developed a partial CE for spherical CdSe and CdS nanocrystals to obtain Cd/Pb chalcogenide heterostructured Janus particles.\cite{Zhang2015} Applying the CE method on elongated nanorods resulted in fully exchanged but geometry-transformed spherical particles.\cite{Zhang2015} In another work, Lee et al.\ demonstrated a direct morphology-preserving CE from Cd to Pb of CdSe nanorods into axial heterojunction nanorods using Pb oleate as a precursor.\cite{Lee2015} A partial CE was also realized on CdSe tetrapods, so that PbSe tips were formed, when \ce{PbBr2} was used as precursor.\cite{Khoshkhoo2020} Apart from CE, single PbS tipped DRs were obtained by direct growth. However, the PbS tips were rather large such that no emission was measured, and no quantization effects were proved.\cite{Rukenstein2012,Rukenstein2016}

In this work, we report on the synthesis of a complex type of heteronanostructure consisting of a spherical CdSe core embedded in a rod-shaped CdS shell that has a PbS tip attached. These CdSe-dot/CdS-rod/PbS-dot nanoparticles (DRDs) are obtained by starting from CdSe-dot/CdS-rod nanoparticles (DRs) and performing a CE reaction with Pb halides in oleylamine as the precursor. The reaction is optimized with respect to the yield of DRDs, which depends on reactant concentrations, reaction times, and DR geometry. 
The photoluminescence spectra of DRDs exhibit dual emission, with one fluorescence band in the near infrared originating from strongly confined charge carriers in the PbS tip and the second peak in the visible spectral region from the CdSe dot.

\section{Results and Discussion}
\subsubsection{General synthesis route}
\begin{figure}[H]
    \centering
    \includegraphics[scale=0.5]{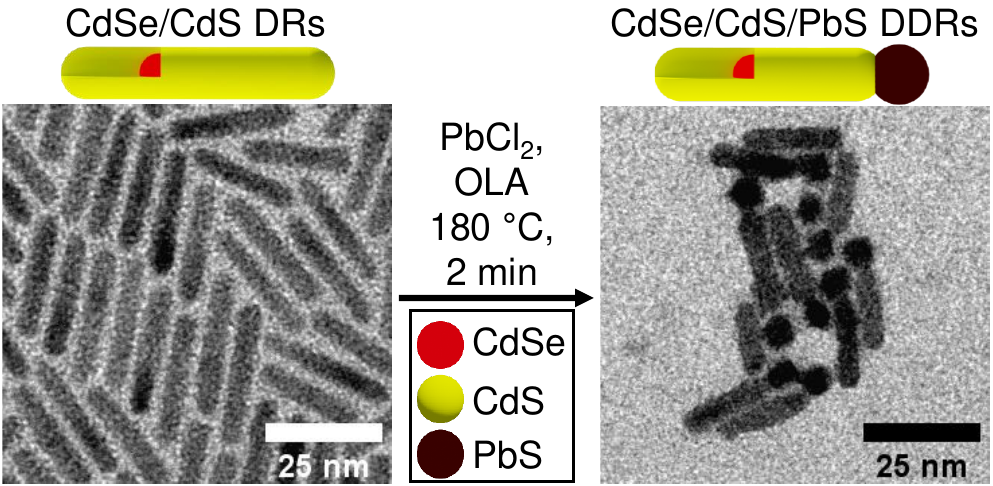}
    \caption{Scheme of a typical reaction. TEM images show the starting material (DRs) and products (DRDs, DRs, QDs).}
    \label{fig:Figure1}
\end{figure}

For the synthesis of CdSe/CdS/PbS DRDs a rather simple approach is employed, as visualized in Fig.\ \ref{fig:Figure1}. Starting point are CdSe/CdS DRs synthesized using the protocol of Carbone and coworkers\cite{Carbone.2007}. These DRs are homogeneous in shape and size, as is revealed by transmission-electron microscopy (TEM) images (Fig.\ \ref{fig:Figure1}a). A dispersion of DRs diluted in 1-octadecene is injected into a solution of lead chloride in oleylamine at a temperature of 180~$^\circ$ C under \ce{N2} atmosphere. After the reaction time (2~min in the case of Fig.\ \ref{fig:Figure1}), the solution is cooled to room temperature and subjected to a standard purification procedure (see Experimental Section). The TEM image in Fig.\ \ref{fig:Figure1}b displays the products obtained by this approach. Here, three different kinds of nanoparticles can be distinguished: (i) elongated nanoparticles with a spherical tip, (ii) elongated structures without a spherical tip, and (iii) spherical nanoparticles. In the following, these structures are referred to as DRDs, DRs, and quantum dots (QDs), respectively.

\subsubsection{Structural analysis}
\begin{figure}[ht]
    \centering
    \includegraphics[scale=0.5]{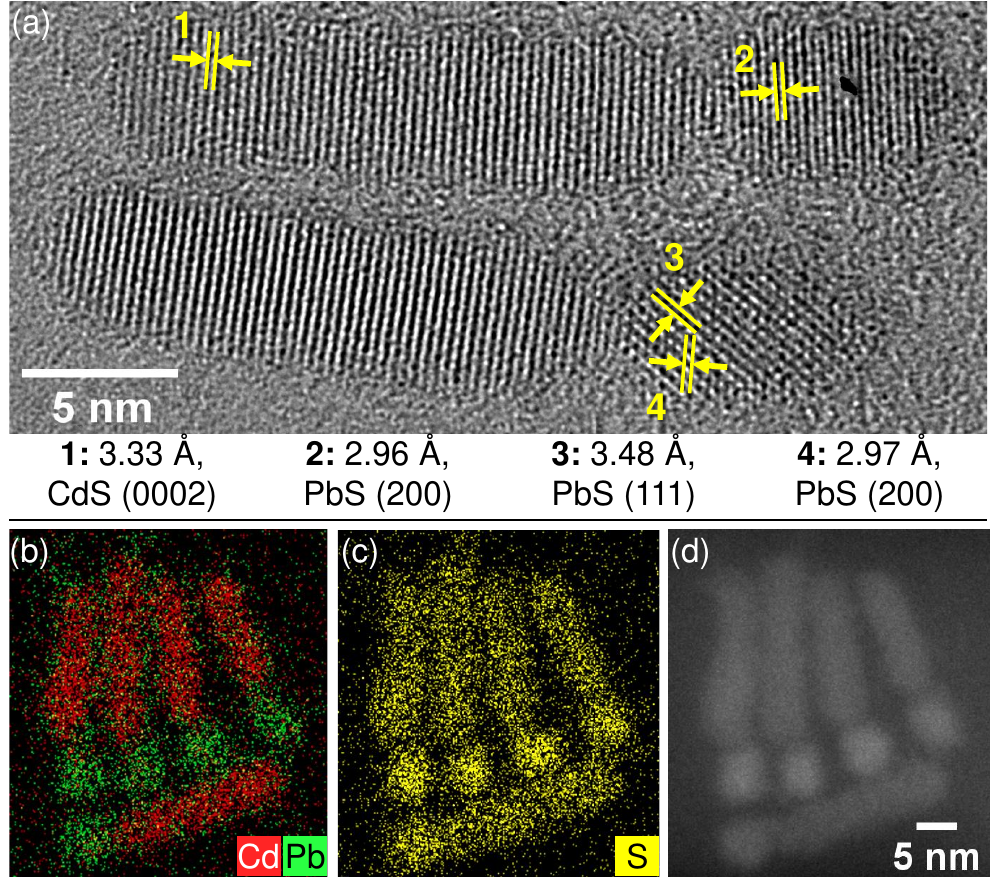}
    \caption{(a) HRTEM  image of two DRDs. The lattice constants were measured and assigned to the lattice planes. (b, c) EDX maps of five DRDs indicating the (b) Cd, Pb and (c) S distribution. (d) HAADF-STEM image of the same DRDs as in (b, c).}
    \label{fig:Figure2}
\end{figure}

Figure \ref{fig:Figure2}a shows a high resolution TEM (HRTEM) image of two representative DRDs. Both nanostructures consist of an elongated part with a more spherical part attached at one of the tips. Between both parts, a constriction occurs. Lattice planes in both parts prove their crystallinity. Analyzing the lattice spacing yields a value of  3.36~\AA{} for the rod-shaped part in the direction of its long axis. This value fits to the (0002)\textsubscript{CdS} planes of hexagonal-phase CdS. The corresponding direction is known to be the growth direction of CdS nanorods.\cite{Talapin2003} The analysis of the more spherical parts yields lattice spacings of 2.96~\AA{} and 3.48~\AA{}, corresponding to the (200)\textsubscript{PbS} and (111)\textsubscript{PbS} planes of cubic PbS, respectively. Which type of lattice plane is visible in HRTEM depends on the orientation with respect to the electron beam. Both DRDs in Fig.\ \ref{fig:Figure2}a reveal that the (0002)\textsubscript{CdS} planes of the rod and the (200)\textsubscript{PbS} planes of the tip form the interface of both parts at the constriction.

Figures \ref{fig:Figure2}b and c show energy dispersive X-ray (EDX) maps of five representative DRDs. 
Obviously, Cd atoms are located in the elongated part of the nanoparticles, Pb atoms are located in the spherical tips, while S atoms occur in both parts. 
A high-angle annular dark-field scanning TEM  (HAADF-STEM) image of the same DRDs shown in Fig.\ \ref{fig:Figure2}d exhibits brighter pixels in the tips than in the rest of the nanoparticles. Since the HAADF-STEM intensity scales with Z\textsuperscript{2} (atomic number Z), this is in line with Pb-containing tips and Cd-containing rods.

In summary, all TEM results presented in Fig.\ \ref{fig:Figure2} confirm a successful synthesis of PbS-tipped CdS rods out of CdSe/CdS DRs. The position of the CdSe dots within the rod-shaped CdS part of the DRDs could not be resolved by TEM, however, their existence can be proved by optical spectroscopy, as will be explained later. The TEM results do not give unambiguous information about the growth mechanism of the PbS tips. Since no extra S precursor was added in the reaction solution, a simple nucleation and growth process at one tip of the DR can be ruled out. If nucleation and growth of PbS tips were to take place, a preceding dissolution process of CdS would be necessary for the provision of a sulfur source, but we have no evidence for this. Hence, we assume a CE process from CdS to PbS  to take place, starting at one tip of the CdSe/CdS DRs.
The preferential exchange at only one of the DR tips can be explained by the non-centrosymmetric wurtzite structure of CdS, which results in different $(0001)$ and $(0001\bar)$ facets, one of which is sulfur-rich and the other Cd-rich.\cite{Talapin2003,Manna2005}
The preferential CE in nanorods has been observed for the Cd-to-Pb exchange in CdSe \cite{Lee2015} and for the Cd-to-Cu exchange in CdS \cite{Sadtler2009}.  For the partial CE of CdSe nanorods, the interface between original and cation-exchanged material was reported to be between \{0001\} planes of the hexagonal and  \{111\} planes of the cubic lattice.\cite{Lee2015} Interestingly, each HRTEM image of our structures that made lattice planes visible showed that the interface is formed between (0002)\textsubscript{CdS} and (200)\textsubscript{PbS} planes. This arrangement has been reported before, for the selective growth of PbSe on CdS nanorods.\cite{Kudera2005}

We assume that the constriction between the CdS rod and the newly formed PbS tip, as observed in TEM images, is promoted by the reorganization of the crystal structure \cite{Son2004} from wurtzite CdS to cubic PbS during CE.  The CE occurs in the full cross-sectional area of the particles perpendicular to the [0001] direction. Hence, the reaction zone, which is the area where cations diffuse in and out of the crystal,\cite{Cho2019} is larger than the particles’ cross-sectional area.

\subsubsection{Yield optimization}

In order to increase the yield of DRDs in the synthesis, which resulted in DRs, DRDs and QDs (see Fig.\ \ref{fig:Figure1}), and to get better insight into the reaction mechanism, two experiments have been performed.

In the first experiment, different amounts of \ce{PbCl2} were used, measured in Equivalents (Eq) related to the number of DRs. For example, 1~kEq \ce{PbCl2} corresponds to 1,000 Pb atoms per DR. Three syntheses were carried out with different precursor amounts, ranging from small ($1.1$~kEq) over medium ($2.0$~kEq) to large ($30$~kEq). For comparison, the estimated number of Cd atoms of a single DR used as the starting material here is approx.\ 18,000. During each synthesis, aliquots of the reaction solution were taken after certain reaction times between 30 and 600 s. The proportion of DRDs, DRs, and QDs in each aliquot has been analyzed using TEM. It should be noted that all structures were counted as DRs if they appeared elongated but without distinct tips. Only those elongated structures where a tip could be clearly identified were counted as DRDs. All other, mainly spherical, structures were counted as QDs.
\begin{figure*}[t]
    \centering
    \includegraphics[scale=0.465]{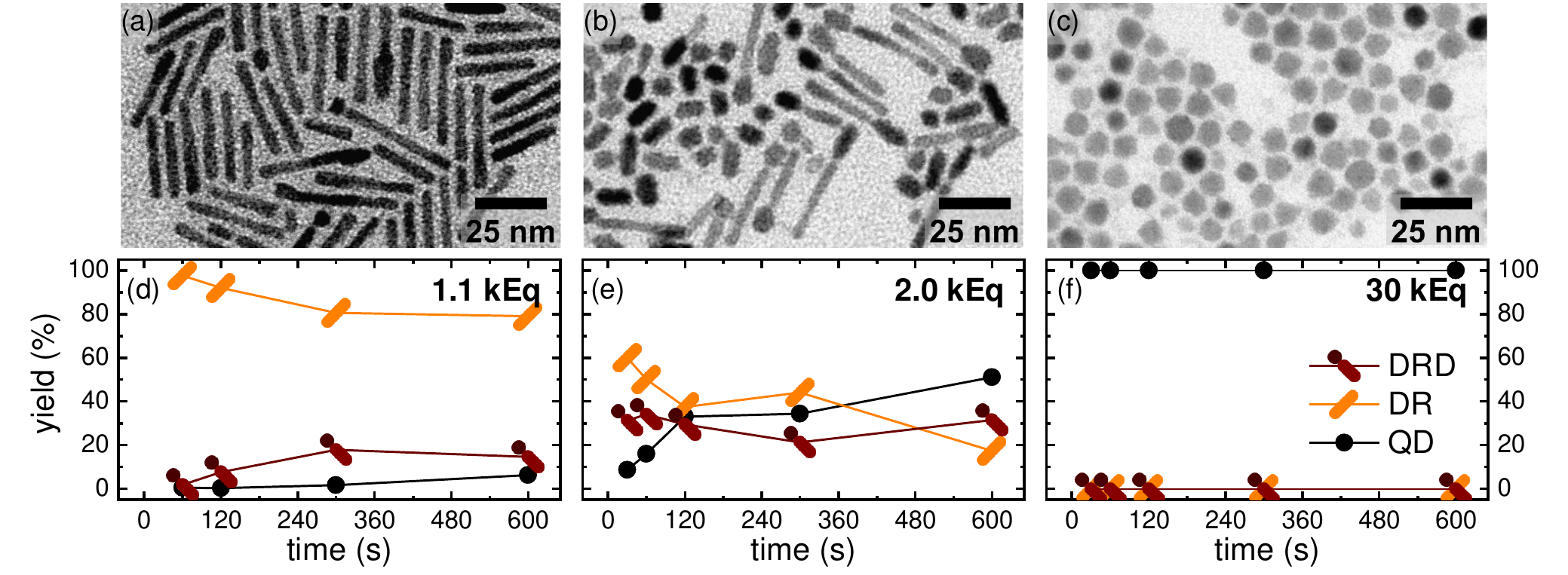}
    \caption{(a-c) Representative TEM images of samples obtained using (a) 1.1 kEq, (b) 2.0 kEq, and (c) 30 kEq of Pb precursor. (d-e) DR, DRD, and QD yields vs.\ reaction time for syntheses using different amounts of Pb precursor.}
    \label{fig:Figure3}
\end{figure*}
Figure \ref{fig:Figure3}a-c shows representative TEM images of the final reaction products of the three syntheses.
Figure \ref{fig:Figure3}d-e shows the yields of DRs, DRDs, and QDs, as measured from TEM images, in dependence of the reaction time for the three different syntheses. 
For the small precursor equivalent, the percentage of DRs in the sample is slowly decreasing over time, reaching about 80\% after 5 min and stays constant after that, suggesting that after 5 min most of the precursor is consumed. The percentage of DRDs initially increases as the yield of DRs is decreasing. A significant proportion of QDs can only be noticed after more than 5 min. Since the increase in the proportion of QDs comes with a slight decrease of DRDs, it might be assigned to the detachment of PbS tip from previously built DRDs.
For the medium precursor equivalent, the percentage of DRs initially decreases rapidly. In the first aliquot, taken after a  reaction time of 30 s, the proportion of DRs has already dropped to about 60\%, while the percentage of DRDs has increased to about 30\%. With increasing reaction time, the percentage of DRs further decreases, similar to the increase in QDs, while the DRD yield remains approximately constant between 20 and 40\%. After a reaction time of about 120 s, DRs, DRDs, and QDs are present in approximately equal quantities.
For the high precursor equivalent, already after a reaction time of only several seconds, all original DRs have turned to QDs. 
Zhang et al.\cite{Zhang2015} reported on a full transformation of elongated CdSe nanorods to spherical PbSe quantum dots in a similar experiment using an excess of \ce{PbCl2} at a temperature of 80 $^\circ$C after a reaction time of 90 min. The increased reaction speed here is a consequence of the elevated temperature of 180 $^\circ$C.

In the second experiment, a set of syntheses was carried out in which the reaction time was set to 120 s but where the precursor amount were varied between 0.8 and 2.7~kEq. The product of each synthesis was again analyzed by TEM. 

\begin{figure*}[b]
    \centering
    \includegraphics[]{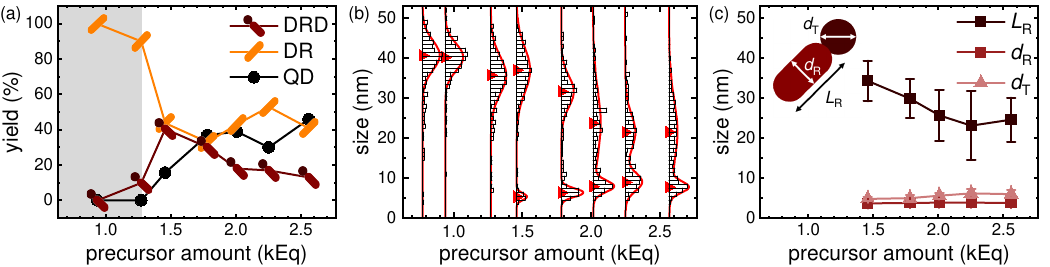}
    \caption{(a) DR, DRD, and QD yields, (b) the largest dimension of each particle, and (c) geometry parameters of synthesized DRDs depicted vs.\ the amount of used Pb precursor.}
    \label{fig:Figure4}
\end{figure*}
Figure \ref{fig:Figure4}a shows the proportions of DRs, DRDs, and QDs for each synthesis in dependence of the used \ce{PbCl2} precursor amount. A minimum amount of 1.3~kEq was needed such that DRDs could be unambiguously discriminated from DR structures. With increasing precursor amount, the proportion of DRs initially decreases rapidly before remaining almost constant in a range between 30 and 50\%. The yield of DRDs initially increases rapidly to a value of around 40\%, before it decreases again at the cost of an increase in the proportion of QDs.

Summarizing both experiments, it can be seen that the equivalent of \ce{PbCl2} precursor used in the synthesis is a crucial and sensitive parameter that determines the DRD yield. By choosing the right equivalent, a 40\% DRD yield could be achieved without further purification of the sample. 

\subsubsection{Reaction mechanism}

From Fig.\ \ref{fig:Figure4}a it can be seen that for high precursor amounts the proportion of DRs stays essentially constant while the yield of DRDs is decreasing. This suggests that a detachment of tips from DRDs takes place, which converts a DRD into a shorter DR and a QD. 
To gain further insight into the reaction mechanism of the DRD synthesis, we further analyzed the TEM images that underlie Fig.\ \ref{fig:Figure4}a. In these images, for each particle ---independent of its classification as a DR, DRD, or QD--- the size in direction of its longest extent has been measured. Figure \ref{fig:Figure4}b shows histograms of lengths as determined for each sample, along with Gaussian fits and corresponding mean values. 

It can be seen that the length of the particles decreases from about 40~nm for the lowest precursor amount to roughly 20~nm for the maximum amount of \ce{PbCl2}. For more than 1.3~kEq, a second maximum in the frequency distribution occurs. This maximum occurring at  smaller lengths represents particles classified as QDs. Their diameter increases with the amount of \ce{PbCl2} added. 

In Fig.\ \ref{fig:Figure4}c the mean geometry parameters of particles classified as DRDs are depicted. The mean length ($L_R$) of the rod-shaped part of the DRDs is decreasing with increasing precursor amount, whereas the diameter of the rod ($d_R$) stays constant. The mean diameter of the tip ($d_T$) is slightly larger than $d_R$. It is slightly increasing with increasing precursor amount but stays smaller than the mean diameter of the QD particles in the respective sample (cf.\ Fig.\ \ref{fig:Figure4}b).

\begin{figure}[b]
    \centering
    \includegraphics[]{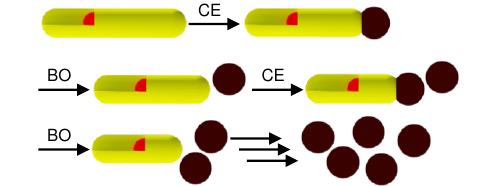}
    \caption{Sketch of the reaction mechanism leading from DRs to DRs, DRDs, and QD by cation exchange (CE) and break-off (BO) events.}
    \label{fig:Figure5}
\end{figure}

From these findings, a reaction mechanism can be proposed, as sketched in Fig.\ \ref{fig:Figure5}. The reaction starts with an anisotropic cation exchange  at the tip of the CdSe/CdS DR induced by the \ce{PbCl2} precursor. During this process, the crystal structure undergoes a transformation from wurtzite CdS to cubic PbS.  With ongoing reaction, the volume of exchanged material increases and a constriction between rod and tip material forms. Eventually, the PbS tip breaks off. The shortened DR then can take part in a further cation-exchange process until the newly formed PbS tip breaks off again. This process can be repeated until the DR is consumed. 
This reaction mechanism makes it clear that the precursor amount and the reaction time are important parameters that determine the respective proportions of DRs, DRDs, and QDs in the reaction product. Their optimization leads to a maximum in DRD yield.

It should be stressed that the DRD yield in above reaction can also be changed by other parameters than reaction time and precursor amount. For example, we also used \ce{PbBr2} and \ce{PbI2} instead of \ce{PbCl2} as precursors for the cation exchange. While the general findings concerning the reaction stayed the same, the different precursors lead to subtle changes in the proportions of DRs, DRDs, and QD in the reaction products, which can be optimized by adjusting again precursor concentration and reaction time. Details on these syntheses can be found in the Supporting Information.
The thickness of DRs that serve as starting point for the synthesis of DRDs is another important parameter that influences the DRD yield. We found that the use of thicker DRs increases the stability of the DRDs, i.e.\, it reduced the break-off  of the tip from the rod, and in increases the DRD yield of a synthesis, as will be shown in the next section.

\subsubsection{Optical properties}

\begin{figure*}[t]
    \centering
    \includegraphics[width=\textwidth]{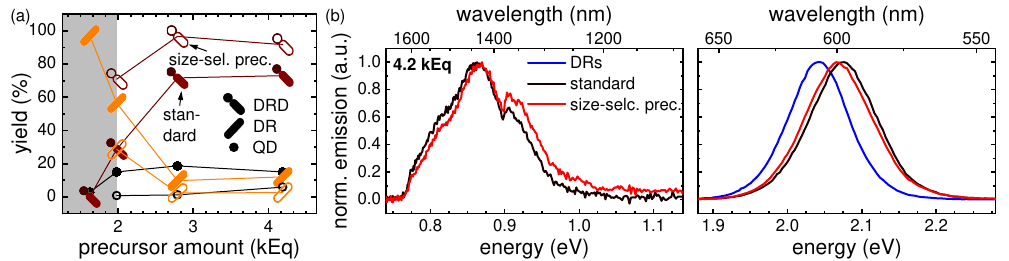}
    \caption{(a) DR, DRD, and QD yields vs.\ precursor amount for a sample purified in a standard way as well as by size-selective precipitation. (b, c) Fluorescence spectra of differently purified DRD samples in (c) the NIR and (d) the visible spectral range. The blue curve in (d) represents the spectrum of the initial DR sample.}
    \label{fig:Figure6}
\end{figure*}

To characterize the optical properties of DRDs, a sample with a maximum DRD yield is needed to avoid spurious signals from DRs and QDs. We reached a DRD yield of $>70 \%$ in a sample that was purified in a standard way (see Experimental Section) by using CdSe/CdS DRs with a diameter of $(4.6 \pm 0.4)$~nm, slightly larger than the diameter of $(3.3 \pm 0.3)$~nm of the DRs used in the syntheses discussed before.
By a size-selective precipitation of the reaction solution (see Experimental Section), the DRD yield was increased to $>95 \%$. This is shown in Figure \ref{fig:Figure6}a, which directly compares the proportion of DR, DRDs, and QDs for a 120~s synthesis in dependence of the \ce{PbCl2} precursor equivalent, for both purification processes. Figure \ref{fig:Figure6}b shows normalized photoluminescence (PL) spectra in the near-infrared (NIR) and the visible spectral region of the samples obtained with a precursor kilo equivalent of 4.2. Here, the excitation laser wavelength was set to 460~nm. The samples show two emission peaks, the NIR emission around 0.86~eV (1440 nm) of the PbS tips of the DRDs and the emission around 2.07~eV (599 nm) of the CdSe cores embedded in the elongated CdS shell. The spectra of the samples treated without and with size-selective precipitation are nearly the same. The slight blue shift in the NIR for the latter sample can be explained by the fact that QDs detached from former DRDs are larger than the mean size of PbS tip when attached to DRDs (cf.\ Figures \ref{fig:Figure4}b and c). The size-selective precipitation removes these larger QDs, which exhibit a red-shifted PL emission due to smaller charge-carrier confinement, thus leaving a sample with a blue-shifted PL emission behind. 
The emission wavelength of the PbS tips correspond to the reported emission wavelength of 5.9-nm-sized PbS nanocrystals synthesized by a hot-injection method.\cite{Caram2016}. The PL linewidth of the PbS tips is about 125~meV, only slightly larger than reported for the PbS QDs,\cite{Caram2016} indicating a slightly broader size distribution of the PbS tips of the DRDs.
The very small red shift of the size-selectively purified sample in the visible range can be reasoned by the removal of shorter DRDs and DRs, so that the remaining sample generally exhibit less charge-carrier confinement.

\section{Conclusion}
We have developed a synthesis for heterostructure nanocrystals consisting of a spherical CdSe core within a rod-shaped CdS shell, capped by a PbS tip. These CdSe-dot/CdS-rod/PbS-dot nanocrystals (DRDs) are synthesized through a cation exchange (CE) method starting from CdSe/CdS-DRs in the presence of Pb halides as precursors in oleylamine. Detailed HRTEM analysis confirmed the successful formation of these structures, showing clear distinctions in lattice structures and elemental compositions (Cd, Pb, S) across the different sections of the nanoparticles. The interface between CdS and PbS was built between the (0002) lattice planes of wurtzite CdS and the (200) lattice planes of rock salt PbS.

Besides DRDs, the synthesis also yielded DRs and QDs as by-products. To optimize the DRD yield, the synthesis parameters, including precursor amounts and reaction times, were varied. It was found that, starting with rather thin DRs (3.3 nm in diameter), a  maximum of 40\% DRD yield could be achieved without sophisticated purification processes by carefully adjusting the synthesis parameters. Based on these findings and on measurements of the nanocrystals' geometry, a mechanism for the reaction could be proposed. At first, the PbS tip is formed by partial CE and crystal reorganization. The tip evolves to a spherical particle that is attached to the DR. Eventually it breaks off, such that a shorter DR and a QD is formed. The (shorter) DR can undergo multiple of these processes. This mechanism also explains the complete conversion of DRs to QDs in the presence of excess Pb precursor. In further experiments, the DRD yield could be increased to over 95\% by using thicker DRs and size-selective precipitation. 

Finally, optical investigations were carried out using these samples. The PL characteristics of the DRDs revealed dual emission peaks: one in the NIR from the PbS tip, indicating strong confinement of charge carriers, and another in the visible region from the CdSe core.

This work advances the potential for precise control over the synthesis of CdSe/CdS/PbS DRD nanocrystals, enabling tailored optical properties for potential applications in optoelectronics and photonics.

\section{Experimental section}

\subsection{Chemicals}

All chemicals were used as received. 
Lead(II) chloride (\ce{PbCl2}, 99.999\%),
1-octadecene (90\%),
oleylamine (70\%),
sulfur powder (S, 99.98\%) and
trioctylphosphine oxide (TOPO, 99\%)
were purchased from Sigma-Aldrich.
Selenium (Se, 99.5\% powder mesh 200)
was purchased from Acros Organics.
Trioctylphosphine (TOP, 97\%) was purchased from ABCR.
Cadmium(II) oxide (CdO, 99.9999\%)
was purchased from ChemPur.
Hexylphosphonic acid (HPA, >99\%) and
octadecylphosphonic acid (ODPA, >99\%) and
were purchased from PCI Synthesis.
Isopropanol (99.7\%)
was purchased from VWR.
Toluene ($\geqq$99.8\%)
and methanol (>= 99.8\%)
were purchased from fischer chemical.
Tetrachloroethylene (TCE, 99+\%)
was purchased from thermo scientific.

\subsection{Synthesis of CdSe/CdS DRs}
CdSe/CdS DRs were synthesized in a two-step synthesis for a seeded-growth approach from a modified previously published method.\cite{Carbone.2007}\\
First, TOPO (3.00~g, 7.76~mmol), ODPA (280~mg, 0.837~mmol) and CdO (60.1~mg, 0.468~mmol) were prepared and evacuated for 1~h at 150~$^\circ$C. Under nitrogen atmosphere the temperature was set to 310~$^\circ$C and hold until a clear solution was obtained. The clear solution was heated to 390~$^\circ$C and TOP (1.8~mL, 4.0~\textmu mol) was added. The temperature was allowed to recover before a solution of selenium in TOP (0.37~mL, 2~\textsc{M}, 0.74~mmol) was added rapidly. After hot injection, the reaction was immediately cooled to room temperature by an air stream.\\
The crude product was purified by the standard purification (cf. section standard purification), before the product was stored in TOP (3.5~mL) under inert conditions.

Then, TOPO (3.00~g, 7.76~mmol), ODPA (290~mg, 0.867~mmol), HPA (80.0~mg, 0.481~mmol) and CdO (86.0~mg, 0.700~mmol) were prepared and evacuated for 1~h at 150~$^\circ$C. Under nitrogen atmosphere the temperature was set to 310~$^\circ$C and hold until a clear solution was obtained. The clear solution was heated to 350~$^\circ$C and TOP (1.8~mL, 4.0~\textmu mol) was added. The temperature was allowed to recover before a solution of selenium in TOP (1.9~mL, 2~\textsc{M}, 3.9~mmol) with CdSe QDs in TOP (153~\textmu L, 0.343~mmol) was added rapidly. After 8~min, the reaction was cooled to room temperature by an air stream.\\
The crude product was purified by the standard purification (cf. section standard purfication), before the product was stored in toluene (2.5~mL).

\subsection{Synthesis of CdSe/CdS/PbS DRDs}

Under vacuum, \ce{PbCl2} (12.06~mg, 43.37~\textmu mol, 4.19~kEq) in oleylamine (5.00~mL, 70~\%) were stirred for 30~min at 140~$^\circ$C. The atmosphere was switched to nitrogen and the reaction mixture was heated to 180~$^\circ$C for 1~h. CdSe/CdS DR/toluene solution (200~\textmu L, 51.7~\textmu mol/L, 0.0103~\textmu mol, 1~Eq) was diluted with 1-octadecene (0.8~mL) and added rapidly. After 2~min, the reaction was cooled to room temperature by a water bath.\\
The crude product was purified either in a standard way or by size-selective precipitation (cf. following sections), before the product was stored either in toluene (2.5~mL) or in TCE (2.5~mL).

\subsection{Standard purification}

The crude product was purified by diluting with toluene (5~mL), precipitating with methanol (10~mL), centrifuging (16,095~RCF) for 10~min and redissolving in toluene (5~mL) or the storing solvent. The purification was done three times. 

\subsection{Purification by size-selective precipitation}

For purification by size-selective precipitation, the crude product was diluted to 20 mL with toluene. One half of the solution was purified in the standard way for reference. The other half was first completely precipitated (centrifuging (16,095~RCF) for 10~min) with isopropanol (10~mL) to remove the first excess chemicals. The precipitate was solved in toluene (10~mL). To the solution, isopropanol (1-4~mL) was stepwise added and after each addition the mixture was centrifuged at 986~RCF for 10~min. 
As soon as a precipitate was observed, the supernatant was removed. Here, the coloration of the supernatant indicates successful size selectivity. The precipitate was then redissolved in toluene (10 ml) and the procedure was carefully repeated starting with the addition of isopropanol. The process was terminated after three successful precipitation steps. Finally, the precipitate was redissolved in either toluene (2.5~mL) or  TCE (2.5~mL).

\subsection{Characterization}

Concentrations of nanoparticles in solutions (CdSe QDs and CdSe/CdS DRs) were determined following the work of Yu and coworkers.\cite{Yu2003} In the determination of CdSe/CdS DR concentration, the shift of the CdSe absorption maximum due to the grown CdS shell was neglected and assumed to be a systematic error. Absorption spectra were recorded on a Varian Cary 5000 UV/vis/
NIR spectrometer in the wavelength range between 300 and 800 nm using quartz cuvettes with a 10 mm path length. The preparation was carried out by defined and systematic dilution of the nanoparticle solutions in toluene.

Transmission electron microscopy (TEM) images were obtained using a JEOL JEM 1011 at an acceleration voltage of 100 kV. High-resolution transmission electron microscopy (HRTEM) and scanning transmission electron microscopy (STEM) images were obtained using a double-corrected (CESCOR and CETCOR, CEOS) JEOL JEM 2200FS microscope with an in-column image filter (Ω-type), a high-angle annular dark-field (HAADF) detector, and a Gatan 4K UltraScan 1000 camera at an accelerating voltage of 200 kV. Energy-dispersive X-ray spectroscopy (EDS) elemental maps were obtained using a JEOL JED-2300 analysis station with a 100 mm\textsuperscript{2} silicon drift detector. For the measurements, the samples were drop-casted on carbon-coated copper TEM grids with 400 meshes (titanium grids with 400 meshes were used for the EDS mappings). For the yield determination and the evaluation of the size distribution of the different nanostructures, ImageJ software was used. For yield determination, a minimum of 300 structures were counted for each sample in different TEM images of different areas of the TEM grids. For size determination, a total of 200 structures were counted for each sample in different TEM images of different areas of the TEM grids. For the determination of lattice spacings, ImageJ and Gwyddion software was used.

The emission spectra were measured with a homebuilt confocal microscope. A pulsed laser diode (Picoquant, LDH-440-P-C-440) with 440~nm emission wavelength was focused by a microscope objective (Mitutoyo NIR APO 100XHR, NA.$ =0.7$)) onto the sample. The emitted and scattered light from the sample was collected with the same objective, filtered by a 532 nm long pass filter and guided into the detection system. The fluorescence light was reflected onto a dichroic beamsplitter (shortpass 950~nm) splitting the light into an NIR beam and a VIS beam. The NIR beam was guided to a spectrometer (Princeton Instruments, Acton SP2300i with  30 gr/mm grating and 1200~nm blaze wavelength) equipped with an InGaAs CCD camera (Princeton Instruments, NIRvana 640), while the VIS beam was guided to a second spectrometer (Andor Shamrock 303 with 50 gr/mm grating and 600~nm blaze wavelength) equipped with a Si EMCCD camera (Andor iXon 897).

\section{Supporting Information}
Supporting Information: Results and Discussion about Effects of Halides (PDF)

\section{Acknowledgment}
We thank the TEM service unit of the University of Hamburg (especially Stefan Werner and Andrea Köppen) for TEM and EDX measurements, 
Sebastian Hentschel for data processing of HRTEM images, and
Luis-Felipe Mochalski for taking part in early stages of the project.

This work was 
funded by the Deutsche
Forschungsgemeinschaft
(DFG, German Research Foundation) via GRK2536 NANOHYBRID, project number 406076438.


\bibliography{bib}

\end{document}